# Formation of ordered arrays of Ag nanowires and nanodots on Si(557) surface

R. A. Zhachuk, S. A. Teys, A. E. Dolbak, and B. Z. Olshanetsky[*]

*Institute of Semiconductor Physics, Russian Academy of Sciences,*
*Siberian Branch, Pr. Lavrentyeva 13, Novosibirsk 630090, Russian Federation*

**Abstract**

Formation of Ag nanostructures on the Si(557) surface containing regular steps of three bilayer height have been studied by scanning tunneling microscopy, low energy electron diffraction and Auger electron spectroscopy at room temperature. It is found that the ordered arrays of nanodots and nanowires of Ag can be formed on this surface. It was shown that a sample exposure in the vacuum before Ag growth affects the shape of the forming Ag islands. This effect is caused by oxygen adsorption on the silicon surface from the residual atmosphere in the vacuum chamber. When Ag is deposited on the clean silicon surface the islands, overlapping several (111) neighboring terraces, form. The arrays of silver nanowires elongated along steps and silver nanodots, arranged in lines parallel to the steps, can be formed on the Si(557) surface depending on the amount of adsorbed oxygen.

[*]Corresponding author. Tel.: +7-3832-333-286; fax: +7-3832-333-502.
*E-mail address:* olshan@isp.nsc.ru (B.Z. Olshanetsky).

## 1. Introduction

Studies of the formation of low-dimensional structures attract considerable interest because the spatial confinement of electrons in these structures results in discrete quantum states. This phenomenon is interesting not only from a fundamental point of view, but can also lead to the development of new devices in the future. Self-organized low-dimensional structures are of particular interest. However, at present there are a limited number of systems with the parameters suitable for the observation of quantum size effects. Most of the published work on self-organized nanostructures is devoted to the nanodots of semiconductors [1, 2]. Few papers are concerned with self-organized nanowires [3,4]. The number of papers concerned with the formation of metallic nanowires and nanodots is even less [5, 6].

A possible way of forming self-organized nanostructures is the employment of vicinal surfaces with regular atomic steps as templates [3-7]. Vicinal Si(111) surfaces inclined towards the [$\bar{1}\bar{1}2$] direction contain one and three bilayer high steps at temperatures below 870°C [8-10]. The share of triple steps increases with an increase of the inclination angle of the vicinal surface [10]. The Si(557) surface is inclined relative to the (111) plane at 9.45° towards the [$\bar{1}\bar{1}2$] direction and it contains only triple steps. The period of the staircase of triple steps on this surface is 5.73 nm. Therefore, the Si(557) surface is promising for the formation of the ordered arrays of nanostructures [11]. The aim of the present work was to explore a possibility of the formation of silver nanowires and nanodots on the Si(557) surface. The study was carried out in an ultra-high vacuum by scanning tunneling microscopy (STM), low energy electron diffraction (LEED) and Auger electron spectroscopy (AES).

## 2. Experimental

The experiments were performed in two separate ultra high vacuum systems, one being equipped with a STM (OMICRON) and the second one with a LEED and an AES (Riber). All STM images were recorded in the constant-current mode using an electrochemically etched tungsten tip. The base pressure in the STM chamber was $9 \times 10^{-11}$ Torr and that in the LEED and AES chamber was $1.6 \times 10^{-10}$ Torr. The pressure of the residual atmosphere in the STM chamber could be raised up to $5 \times 10^{-10}$ Torr by slightly opening the gate to the lock chamber.

Si(111) and Si(557) substrates were used. Samples were resistively heated by a direct current. The current direction was parallel to the steps on the Si(557) samples to avoid from the electromigration induced step bunching. The temperature of a sample was monitored by an optical-disappearing filament pyrometer. Clean surfaces were prepared by carefully degassing a sample at 600°C followed by a short flash to 1250°C to desorb the native oxide. Si(557) samples were prepared following the procedure described in Ref. [11]. After flashing at 1250°C the temperature of the Si(557) sample was lowered quickly to 1060°C, and the sample was sustained at this temperature for 30 seconds. Then the sample was quenched to 830°C, annealed at this temperature for 20 min, and cooled down to room temperature (*RT*) during 20 min.

As a source of Ag atoms we used pieces of Ag welded to a tantalum ribbon. The ribbon was heated by electric current. Ag was deposited on a Si substrate at *RT*. We calibrated the Ag flux in the LEED-AES chamber, using the break in the dependence of the Auger signal of Ag versus silver coverage on a Si(111) surface obtained at 400°C. This was used as the reference point [12]. The break is due to the saturation of the Si(111)-$\sqrt{3}\times\sqrt{3}$-Ag surface structure and to the transition to the formation of 3-dimensional islands at Stranski-Krastanov growth mode. According to the honeycomb-chained triangle model [13], the saturation coverage of the Si(111)-$\sqrt{3}\times\sqrt{3}$-Ag surface structure is 1.0 ML (ML is monolayer). That is, the Si(111)-$\sqrt{3}\times\sqrt{3}$-Ag surface contains one Ag atom per each Si surface atom. In the STM chamber the Ag flux was calibrated measuring the share of a surface area with a Si(111)-$\sqrt{3}\times\sqrt{3}$-Ag structure after depositing a submonolayer amount of silver on a Si(111) sample, which was kept at 400°C. In our experiments the deposition rate was about 1.0 ML per hour.

The surface composition was determined by AES, using Auger Si LVV (92 eV), Ag MNN (351 eV) and O KLL (503 eV) electron peaks. The sensitivity factors were taken from Ref. [14]. To calculate the surface concentration of oxygen, we used a model of uniformly distributed O submonolayer on a Si substrate [15]. The detection limit of oxygen was about 0.02 ML.

## 3. Results

During the initial stage of silver adsorption on the Si(111) surface at *RT* a dilute silver layer forms [13, 16, 17]. The dilute layer can be seen in the inset of Fig. 1(a). This layer shows no ordered structure as it does not bring any new diffraction spots in LEED patterns and causes only the redistribution of spot intensities from the Si(111)-7×7 structure. This diffraction pattern corresponds to the so called δ-7×7 structure [13]. The formation of the dilute layer is completed when the silver coverage is approximately 0.5 ML. After that, the 2-dimensional epitaxial Ag islands form (Fig. 1(a)).

Figure 1(b) shows a diffraction pattern from a Si(111) surface after the adsorption of 6.0 ML of silver at *RT*. It follows from this pattern that the flat tops of the Ag islands have the (111) orientation. The diffraction spots from the Ag islands are azimuthally broadened and have an angular width of about 6°. This is due to random rotational disorder of silver islands relative to the crystallographic orientation of the Si(111) substrate [13, 18].

Figure 2(a) shows an STM image of a Si(557) surface after the adsorption of 4.0 ML of silver at *RT*, immediately after the cleaning procedure. One can see that the array of randomly scattered wide Ag islands have formed on the silicon surface. The size of the silver islands varies a lot. The islands overlap several Si(111) terraces. It was established that in the coverage range from 2.0 to 6.0 ML of silver the islands density is kept constant while the average size of islands increases.

The shape of Ag islands being formed depends drastically on the preliminary exposure of the sample in a vacuum. Figure 2(b) shows an STM image of the array of Ag nanowires formed on the Si(557) surface after its exposure in the vacuum chamber at a pressure of $5.0 \times 10^{-10}$ Torr for 15 minutes, followed by the adsorption of 4.0 of Ag at *RT*. The width of the majority of Ag nanowires is close to that of the Si(111) terraces and amounts to 5 nm. In our experiments sparse wide Ag islands overlaping two or more Si(111) terraces also formed on this surface. The average thickness of the nanowires is equal to that of the four Ag(111) monolayers, i.e. it is 0.944 nm. We could obtain nanowires with a length of up to 150 nm at a Ag coverage of 6.0 ML. However, the increase of the Ag coverage causes an abrupt increase in the number of wide Ag islands.

Sample exposure at a pressure of $5.0 \times 10^{-10}$ Torr for two hours before Ag growth leads to the formation of arrays of Ag nanodots. Figure 2(c) shows the STM image of a Si(557) surface after the adsorption of 2.0 ML of silver at *RT*. The formed nanodots are located on the Si(111) terraces and are arranged in lines parallel to the triple steps. The lateral sizes of nanodots are limited by the width of the Si(111) terrace, so the size distribution is relatively narrow. The average height of the nanodots shown in Fig. 2(c) is equal to the height of three monolayers of Ag(111), i.e. it is 0.708 nm.

Figure 2(d) shows the STM image of Ag nanodots formed on the Si(111) surface under conditions similar to those at which the nanodots on the Si(557) surface were formed (see Fig. 2(c)). One can see that in the absence of triple steps the nanodots are randomly distributed on the surface.

Apparently, the change of the shape of Ag islands is caused by the adsorption of impurities on the silicon surface from a residual atmosphere in the vacuum chamber. The only impurity we found by AES on Si(111) and Si(557) surfaces after the exposure at a pressure of $1.6 \times 10^{-10}$ Torr for 24 hours was oxygen. The oxygen coverage on Si(111) and Si(557) surfaces after the exposure was 0.06 ML and 0.095 ML, respectively. The Si(557) surface consists of (111) terraces and triple steps having, according to different data, either (112) or (113) orientation [11, 19]. Thus, the Si(557) surface area in contact with the residual gases is only 1.1 times bigger than that of the Si(111) surface. This can not be the cause of the difference between the values of oxygen coverage on the Si(111) and Si(557) surfaces. The most plausible explanation for this difference is the preferential adsorption of oxygen atoms on the triple step edges, probably, because the density of the dangling bonds on the step edges is higher than that on Si(111) terraces.

As long as the oxygen concentration is low, the amount of adsorbed oxygen on the silicon surface is proportional to the exposition $pt$. Here $p$ is the pressure of residual atmosphere in the vacuum chamber and $t$ is the time of exposure. Therefore, it is possible to estimate the oxygen concentrations on the Si(557) surface after different exposures. The estimated value of the oxygen concentration on the surface with nanowires (Fig. 2(b)) is 0.003 ML and that for the surface with nanodots (Fig. 2(c)) is 0.025 ML.

Thus, adsorbed oxygen on the silicon surface acts like a surfactant affecting the shape of islands during the Ag growth [20]. The possible mechanism of this effect can be described as follows. Atoms of the adsorbed oxygen saturate the dangling bonds on the silicon surface and thus change the surface free energy. As a result Ag does not wet the surface areas covered by oxygen. Ag islands could be formed only within the areas which did not contain oxygen. At low expositions oxygen atoms are incorporated into the edges of the triple steps. Therefore, at low coverages Ag forms nanowires on (111) terraces of the Si(557) surface. With the increase of an exposition, the oxygen coverage on (111) terraces increases and this results in the formation of nanodots.

Figure 3(a) shows the diffraction pattern from the Si(557) surface covered by 2.0 ML of silver at *RT* after the adsorption of 0.012 ML of oxygen. On the surface the Ag nanowires and nanodots, similar to those shown in Figs. 2(b) and 2(c), are present. The mutual positions of the diffraction spots from the Ag islands indicate that the flat tops of the islands have the (111) orientation. Moreover, the profile of the (00) spot shows that the positions of the (00) spot from the flat tops of the Ag islands and those from the Si(111) terraces coincide (Fig. 3(c)). Therefore, the flat tops of the silver islands are parallel to the Si(111) plane. It follows from Fig. 3(a) that there is no azimuthal broadening of the (01) and (10) spots from Ag islands, as in the case of Ag adsorption on a flat Si(111) surface (Fig. 1(b)) and therefore there is no random rotational disorder of Ag islands on the Si(557) surface.

Figure 3(b) shows the diffraction pattern from the Si(557) surface, which is covered by 5.0 ML of Ag at *RT* after the adsorption of 0.001 ML of oxygen. The Ag nanowires similar to those presented in Fig. 2(b), as well as wide Ag islands, shown in Fig. 2(a), have formed on the surface. The profiles of the diffraction spots in Fig. 3(c) show that there are two sets of spots from the tops of the silver islands: (00), (10) and (00)', (10)'. Apparently, the first set of diffraction spots comes from the Ag nanowires and nanodots, whereas the second set comes from the wide Ag islands. The position of the second set of spots is shifted relative to the first one by $d$ = 10% SBZ. SBZ is the Surface Brillouin Zone which corresponds to the distance between the integral order spots of Si(111). The shift of the diffraction spots is due to the small inclination of the flat tops of wide silver islands relative to the Si(111) terraces in the $[\bar{1}\bar{1}2]$ direction. The inclination angle is $\alpha = \arcsin(k_\parallel/k)$, where $k$ is the length of the wave vector of primary electrons, and $k_\parallel$ is the component of the wave vector of reflected electrons, which is parallel to the Si(111) terraces for the (00)' spot. For the primary electrons with the energy of 45 eV the wave vector is $k = 34$ nm$^{-1}$. The component $k_\parallel = da^*/100 = 1.9$ nm$^{-1}$, where $a^* = 2\pi/(0.384\sin 60°)$ nm$^{-1}$ is the unit translation vector of the unreconstructed Si(111) surface in reciprocal space. Therefore, the flat tops of the Ag islands are inclined relative to the Si(111) terraces by the angle $\alpha \approx 3.2°$.

The inclination of the flat tops of wide silver islands relative to the flat tops of nanowires and nanodots was confirmed by STM experiments (Fig. 4(a)). It was found that the angle between the (111) top of an Ag island and its side face oriented in the step-down direction is about 55°. Therefore, the side plane has the (100) orientation.

To illustrate the inclination of wide silver islands relative to the Si(111) terraces, a model similar to that proposed for Pb film on Si(779) and Si(557) surfaces [21] can be used. The (111) interplanar distances in Si and Ag are 0.314 nm and 0.236 nm, respectively. Thus, the difference between the heights of a triple Si step and a triple Ag(111) layer is 0.234 nm. This is the reason for the inclination of wide Ag islands relative to Si(111) terraces, as it is shown schematically in Fig. 4(b). The calculated inclination angle of wide Ag islands is $\alpha = \arcsin(0.234/5.73) = 2.3°$, which is close to the experimentally measured value.

The authors of Ref. [22] observed photoelectron peaks from an epitaxial Ag film with a thickness of 5 nm on Si(111) at *RT*. These peaks are associated with spatial confinement of electrons in the direction perpendicular to the surface of the film which results in discrete quantum states. Silver nanostructures produced in present work have similar dimensions and therefore they may also exhibit quantum size effects.

## 4. Conclusions

Formation of Ag nanostructures has been studied on the Si(557) surface containing regular three bilayer-high steps. It is established that the ordered arrays of Ag nanodots and nanowires can be formed on this surface. It is found that the adsorption of a small amount of oxygen from the residual atmosphere strongly affects the shape of growing silver islands. When silver is deposited on a clean Si(557) surface at *RT* it forms wide silver islands which overlap several neighboring terraces and are randomly scattered over the Si surface. The

7arrays of epitaxial Ag nanowires, elongated along triple steps, can be formed on the Si surface with an oxygen coverage of 0.003 ML. At an oxygen coverage of 0.025 ML silver forms arrays of epitaxial nanodots with lateral sizes comparable with the width of Si(111) terraces. Nanodots are located on (111) terraces and are arranged in lines parallel to the triple steps.

All Ag islands have flat tops of the (111) orientation. The side planes faced in the step-down direction have the (100) orientation. The Ag islands grown on flat Si(111) surfaces are azimuthally disordered, while the islands grown on Si(557) surfaces show no rotational disorder. Ag nanowires and nanodots have flat tops parallel to the terrace plane. Flat tops of the wide silver islands ovelapping several Si(111) terraces are inclined at about 3° in the step-down direction relative to the (111) terraces. The inclination of the flat tops of wide silver islands is explained within the framework of a simple model based on the misfit of Ag and Si interplanar (111) distances.

## Acknowledgements

This work was supported by the Russian Foundation for Basic Research and by the Ministry of Science, Industry and Technology of Russia (Russian Federal Program).

## References

[1] O. G. Schmidt, S. Kiravittaya, Y. Nakamura, H. Heidemeyer, R. Songmuang, C. Müller, N. Y. Jin-Phillipp, K. Eberl, H. Wawra, S. Christiansen, H. Gräbeldinger, and H. Schweizer, Surf. Sci. 514 (2002) 10.

[2] O. P. Pchelyakov, Yu. B. Bolkhovityanov, A. V. Dvurechenskii, A. I. Nikiforov, A. I. Yakimov, and B. Voigtländer, Thin Solid Films 367 (2000) 75.

[3] H. Omi and T. Ogino, Phys. Rev. B 59 (1999) 7521.

[4] M. Kawamura, N. Paul, V. Cherepanov, and B. Voigtländer, Phys. Rev. Lett. 91 (2003) 096102.

[5] M. Jalochowski and E. Bauer, Surf. Sci. 480 (2001) 109.

[6] I. K. Robinson, P. A. Bennett, and F. J. Himpsel, Phys. Rev. Lett. 88 (2002) 096104.

[7] R. A. Zhachuk, S. A. Tiis, and B. Z. Ol'shanetskii, JETP Letters 79 (2004) 381.

[8] V. I. Mashanov and B. Z. Ol'shanetskii, JETP Lett. 36 (1982) 355.

[9] B. Z. Olshanetsky and S. A. Teys, Surf. Sci. 230 (1990) 184.

[10] Jian Wei, X.-S. Wang, J. L. Goldberg, N. C. Bartelt, and Ellen D. Williams, Phys. Rev. Lett. 68 (1992) 3885.

[11] A. Kirakosian, R. Bennewitz, J. N. Crain, Th. Fauster, J.-L. Lin, D. Y. Petrovykh, and F. J. Himpsel, Appl. Phys. Lett. 79 (2001) 1608.

[12] G. Le Lay, M. Manneville, and R. Kern, Surf. Sci. 72 (1978) 405.

[13] S. Hasegawa, X. Tong, S. Takeda, N. Sato, and T. Nagao, Prog. Surf. Sci. 60 (1999) 89.

[14] P. W. Palmberg, G. E. Riach, R. E. Weber, and N. C. Mac-Donnald, *Handbook of Auger Electron Spectroscopy* (Phys. Elek. Ind. Inc., Minnesota, 1972).

[15] D. Brigs and M. P. Seach, *Practical Surface Analysis by Auger and X-Ray Photoelectron Spectroscopy,* (John Willey & Sons, Chichester, New York, Brisbane, Toronto, Singapore, 1983).

[16] L. Gavioli, K. R. Kimberlin, M. C. Tringides, J. F. Wendelken, and Z. Zhang, Phys. Rev. Lett. 82 (1999) 129.

[17] L. Huang, S. J. Chey, and J. H. Weaver, Surf. Sci. 416 (1998) L1101.

[18] F. Moresco, M. Rocca, T. Hildebrandt, and M. Henzler, Surf. Sci. 463 (2000) 22.

[19] M. Henzler and R. Zhachuk, Thin Solid Films 428 (2003 129.

[20] C. Y. Fong, M. D. Watson, L. H. Yang, and S. Ciraci, Modelling Simul. Mater. Sci. Eng. 10 (2002) R61.

[21] E. Hoque, A. Petkova, and M. Henzler, Surf. Sci. 515 (2002) 312.

[22] A. L. Wachs, A. P. Shapiro, T. C. Hsieh, and T.-C.Chiang, Phys. Rev. B 33 (1986) 1460.

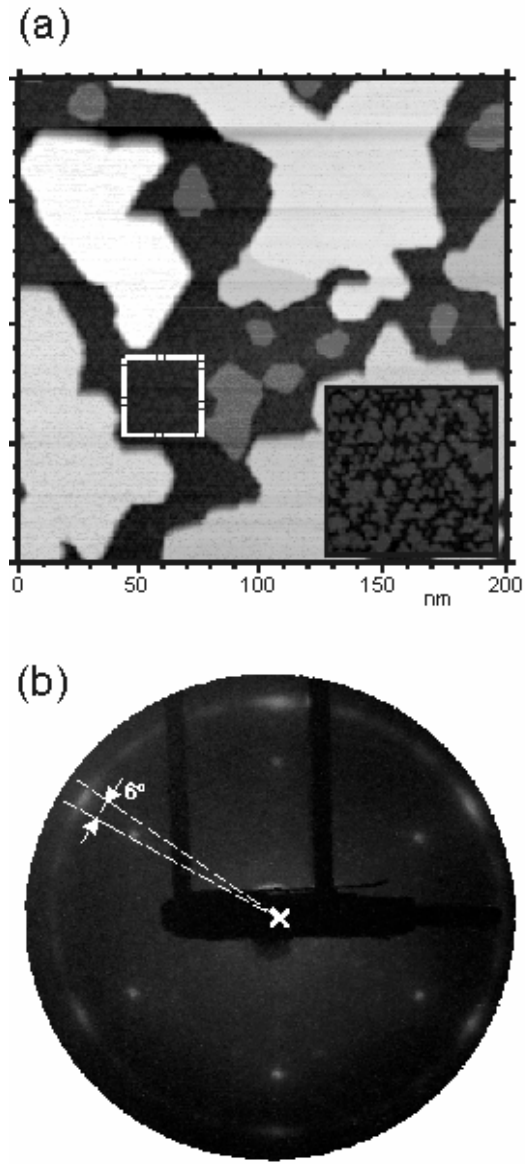

Fig.1. (a) STM image of Si(111) surface after adsorption of 6.0 ML of silver at *RT*; $U$ = +1.5 V, $I$ = 1.52 nA. The inset shows the magnified image of a dilute Ag layer between epitaxial Ag islands. (b) LEED pattern of the Si(111) surface after the adsorption of 6.0 ML of Ag at *RT*. Electron energy is 50 eV.

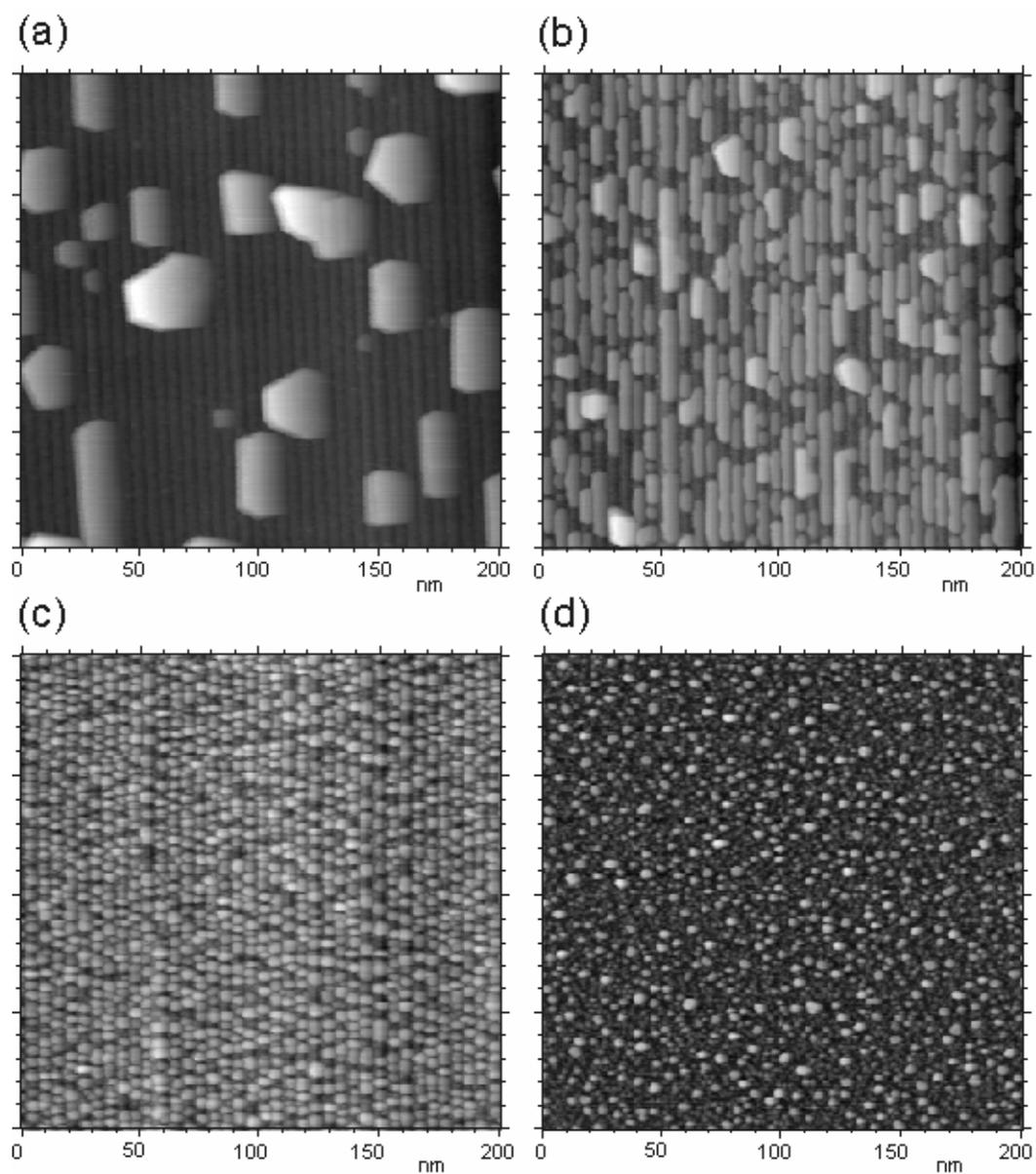

Fig. 2. STM image of Si(557) surface, $U$ = +2.0 V, $I$ = 0.03 nA: (a) Ag coverage is 4.0 ML. (b) Silver coverage is 4.0 ML, oxygen coverage is 0.003 ML. (c) Silver coverage is 2.0 ML, oxygen coverage is 0.025 ML. (d) STM image of Si(111) surface prepared in the same way as Si(557) in Fig. 2(c).

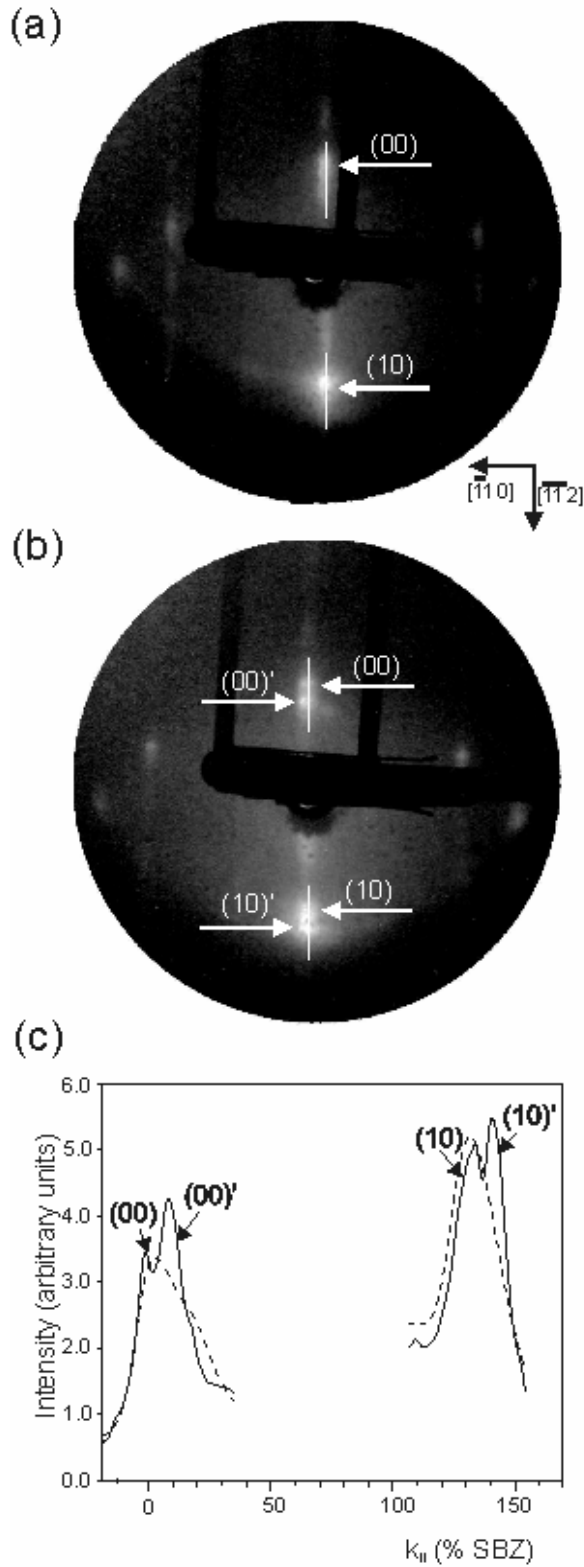

Fig. 3. (a) LEED pattern of Si(557) surface, 45 eV, silver coverage is 2.0 ML, oxygen coverage is 0.012 ML; (b) LEED pattern of Si(557) surface, 45 eV, silver coverage is 5.0 ML, oxygen coverage is 0.001 ML. (c) Profiles of the diffraction spots from Ag islands along the white lines in Figs. 3(a) and 3(b). Dashed lines are the profiles of the (00) and (10) spots in Fig. 3(a), solid lines are the profiles of the (00), (10) and (00)', (10)' spots in Fig. 3(b).

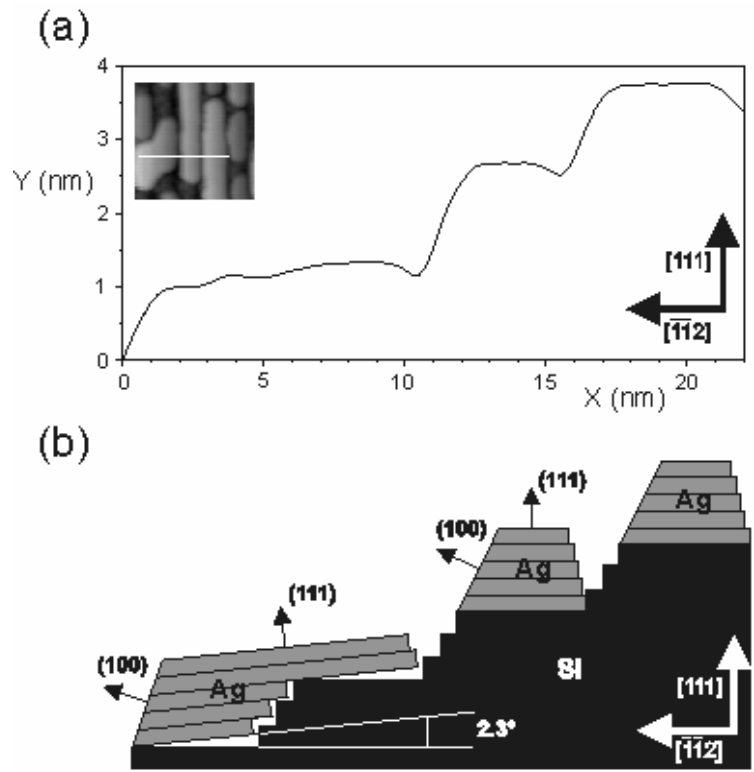

Fig. 4. (a) Profile of Ag nanowires and of a wide Ag island obtained by STM on the Si(557) surface. The inset presents a part of the Si(557) surface with Ag islands. The white line indicates the direction of the profiling. (b) A schematic model of Ag islands on the Si(557) surface.